\documentclass[12pt]{article}
\usepackage{amsmath,amssymb,amsthm,amsxtra,bbm,overpic,bm,epsfig,subfigure}
\usepackage{hyperref}
\usepackage{mathrsfs}
\usepackage{enumitem}
\usepackage{graphicx}
\usepackage{color}
\usepackage{comment}
\usepackage{epstopdf}
\usepackage{float}
\usepackage{cite}
\allowdisplaybreaks[2]

\usepackage{slashed,stmaryrd,multirow}

\def\thefootnote{\fnsymbol{footnote}}

\usepackage{hyperref}
\usepackage{slashed,stmaryrd,orcidlink}

\usepackage{lscape}%
\usepackage{array}
\usepackage{booktabs}%

\begin{document}
	
	\vspace{0.2cm}
	
	\begin{center}
		{\Large\bf Experimental Constraints on Seesaw Parameters in the Wigner-like Parametrization}
	\end{center}
	
	\vspace{0.2cm}
	
	\begin{center}
		{\bf Jihong Huang}~{\orcidlink{0000-0002-5092-7002}}~$^{1,2}$~\footnote{E-mail: huangjh@ihep.ac.cn}
		\\
		\vspace{0.2cm}
		{\small
			$^{1}$Institute of High Energy Physics, Chinese Academy of Sciences, Beijing 100049, China\\
			$^{2}$School of Physical Sciences, University of Chinese Academy of Sciences, Beijing 100049, China}
	\end{center}

	\vspace{0.5cm}
	
	\begin{abstract}
		By introducing three right-handed neutrino singlets, the popular canonical seesaw mechanism is able to simultaneously explain the tiny masses of Majorana neutrinos and the baryon asymmetry of the Universe. In this paper, we provide an explicit calculation in this model with the help of the Wigner-like parametrization. We work in a special ansatz where both ${\bf m}_{\rm D}^\dagger {\bf m}_{\rm D}^{}$ and ${\bf m}_{\rm R}^\dagger {\bf m}_{\rm R}^{}$ are diagonal, with ${\bf m}_{\rm D}^{}$ and ${\bf m}_{\rm R}^{}$ being accordingly the Dirac and Majorana neutrino mass matrices, and $[{\bf m}_{\rm D}^\dagger {\bf m}_{\rm D}^{}, {\bf m}_{\rm R}^\dagger {\bf m}_{\rm R}^{}] = {\bf 0}$ holds. Physical observables can be exactly calculated without any approximation, where three light Majorana neutrino masses $m_i^{}$, leptonic mixing angles $\theta_{ij}^{}$, CP-violating phases $\{\delta,\rho,\sigma\}$, and three rotation angles $\vartheta_i^{}$ describing the hierarchy between electroweak and seesaw scales are chosen as input parameters. For demonstration, we evaluate the branching fractions of the lepton-flavor-violating decays of charged leptons and the CP-violating asymmetries in the resonant thermal leptogenesis. The model parameters are constrained by the latest experimental limits.
	\end{abstract}

	\def\thefootnote{\arabic{footnote}}
	\setcounter{footnote}{0}
	
	\newpage
	
	\section{Introduction}
	
	\label{sec:intro}
	
	The canonical seesaw mechanism can explain the neutrino mass origin in a natural and economical way by extending the Standard Model (SM) with three right-handed neutrino singlets $N_{\rm R}^{}$ ~\cite{Minkowski:1977sc,Yanagida:1979as,Gell-Mann:1979vob,Glashow:1979nm,Mohapatra:1979ia}. In this model, the tiny masses of ordinary Majorana neutrinos originate from the suppression of heavy ones, aligning with the idea of integrating out heavy degrees of freedom to obtain the SM effective field theory~\cite{Weinberg:1979sa}. Furthermore, the accompanied thermal leptogenesis can also explain the observed matter-antimatter asymmetry of the Universe~\cite{Fukugita:1986hr}, which is one of the most important questions for particle physics and cosmology. This renormalizable model can be described by the following Lagrangian:
	\begin{eqnarray}
		\label{eq:type-I-lag}
		{\cal L} = {\cal L}_{\rm SM}^{} + \overline{N_{\rm R}^{}} {\rm i} \slashed{\partial} N_{\rm R}^{} - \left[\overline{\ell_{\rm L}^{}} \widetilde{H} {\bf y}_\nu^{} N_{\rm R}^{} + \frac{1}{2} \overline{N_{\rm R}^{\rm c}} {\bf m}_{\rm R}^{}  N_{\rm R}^{} + {\rm h.c.} \right] \;,
	\end{eqnarray}
	where ${\cal L}^{}_{\rm SM}$ is the Lagrangian of the SM, $\widetilde{H} \equiv {\rm i} \sigma^2_{} H^*$ with $H$ being the SM Higgs doublet, $\ell^{}_{\rm L} = \left(\nu_{\rm L}^{}, l_{\rm L}^{}\right)^{\rm T}$ is the left-handed lepton doublet. With ${\cal C} \equiv {\rm i}\gamma^2 \gamma^0$, we have the charge-conjugate counterpart of the right-handed neutrino singlet $N_{\rm R}^{\rm c} \equiv {\cal C} \overline{N_{\rm R}^{}}^{\rm T}$. In addition, the $3\times 3$ Yukawa coupling matrix ${\bf y}_\nu^{}$ and the symmetric Majorana mass matrix ${\bf m}_{\rm R}^{}$ are denoted in bold. After the spontaneous symmetry breaking (SSB) of the gauge symmetry ${\rm SU}(2)_{\rm L}^{} \otimes {\rm U}(1)_{\rm Y}^{}$, the Higgs field acquires its vacuum expectation value $\left<H\right>=\left(0,v/\sqrt{2}\right)^{\rm T}$ with $v\approx 246~{\rm GeV}$, and we are left with the Dirac neutrino mass matrix ${\bf m}_{\rm D}^{} \equiv {\bf y}_\nu^{} v/\sqrt{2}$.
	
	Although the seesaw mechanism is quite appealing, reasonable approximations are usually required to calculate physical quantities due to the intricate relations among seesaw parameters, which in turn undermines the feasibility of the model. A natural question is whether it is possible to perform a rigorous calculation without any approximation. In this work, we attempt to answer this question with the help of the Wigner-like parametrization~\cite{Wigner:1968,Zhou:2025qwk}. First, we express all model parameters in the flavor basis where the commutation condition $[{\bf m}_{\rm D}^\dagger {\bf m}_{\rm D}^{}, {\bf m}_{\rm R}^\dagger {\bf m}_{\rm R}^{}] = {\bf 0}$ is satisfied. With such a special {\it ansatz}, all physical observables can be calculated {\it explicitly} and {\it rigorously} in terms of the chosen input parameters: three light Majorana neutrino masses $\{m_1^{}, m_2^{}, m_3^{}\}$, three mixing angles $\{\theta_{12}^{}, \theta_{13}^{}, \theta_{23}^{}\}$, three CP-violating phases $\{\delta, \rho, \sigma\}$ and three introduced rotation angles $\left\{\vartheta_1^{}, \vartheta_2^{}, \vartheta_3^{} \right\}$ to describe the huge hierarchy between the electroweak (EW) scale $\Lambda_{\rm EW}^{} \approx 10^2~{\rm GeV}$ and the seesaw scale $\Lambda_{\rm SS}^{} \approx 10^{14}~{\rm GeV}$. Then, as a demonstration, we calculate the branching fractions of the lepton-flavor-violating (LFV) decays of charged leptons $\beta^- \to \alpha^- + \gamma$ for $(\alpha,\beta) = (e,\mu), (e,\tau)$ and $(\mu,\tau)$, and the CP-asymmetries $\varepsilon_{i\alpha}^{}$ in the flavored resonant leptogenesis. Such a {\it bottom-up} approach enables to impose strong constraints on model parameters from experimental measurements and mitigates the degeneracy when evaluating seesaw parameters. 
	
	The remaining parts of this paper are organized as follows. We use the Wigner-like parameterization to express the seesaw model parameters and discuss their properties in Sec.~\ref{sec:seesaw}. With the chosen physical parameters as inputs, the explicit calculations of LFV decay rates and the CP-violating asymmetries in thermal leptogenesis are performed in Sec.~\ref{sec:mutoegamma} and Sec.~\ref{sec:leptogenesis}, respectively, and the experimental constraints of model parameters are thereby determined. Our results and conclusions are summarized in Sec.~\ref{sec:sum}. Finally, for completeness, the exact formulae of CP-violating parameters are listed in Appendix~\ref{app:A}.

	\section{Seesaw Model in Wigner-like Parametrization}
	
	\label{sec:seesaw}

    After the SSB, the mass terms in Eq.~(\ref{eq:type-I-lag}) can be written in a more compact form:
	\begin{eqnarray}
		\label{eq:Lmass}
		{\cal L}_{\rm mass}^{} = -\frac{1}{2} \overline{\begin{pmatrix} \nu_{\rm L}^{} & N_{\rm R}^{\rm c} \end{pmatrix}} 
		\begin{pmatrix}
			{\bf 0} & {\bf m}_{\rm D}^{} \\ {\bf m}_{\rm D}^{\rm T} & {\bf m}_{\rm R}^{} 
		\end{pmatrix} 
		\begin{pmatrix}
			\nu_{\rm L}^{\rm c} \\ N_{\rm R}^{} 
		\end{pmatrix} + {\rm h.c.} \;.
	\end{eqnarray}
	One can diagonalize the $6\times6$ mass matrix by a $6\times6$ unitary matrix ${\cal U}$, i.e.,
	\begin{eqnarray}
		\label{eq:diagonalize}
		\begin{pmatrix}
			{\bf V} & {\bf R} \\ {\bf S} & {\bf U}
		\end{pmatrix}^\dagger \begin{pmatrix}
			{\bf 0} & {\bf m}_{\rm D}^{} \\ {\bf m}_{\rm D}^{\rm T} & {\bf m}_{\rm R}^{} 
		\end{pmatrix} \begin{pmatrix}
			{\bf V} & {\bf R} \\ {\bf S} & {\bf U}
		\end{pmatrix}^* =  \begin{pmatrix}
			\widehat{\bf m} & {\bf 0} \\ {\bf 0} & \widehat{\bf M}
		\end{pmatrix} \;,
	\end{eqnarray}
	where ${\bf V}$, ${\bf R}$, ${\bf S}$ and ${\bf U}$ are all $3\times 3$ matrices satisfying the unitarity conditions ${\bf U}^\dagger {\bf U}+ {\bf R}^\dagger {\bf R} = {\bf V}^\dagger {\bf V} + {\bf S}^\dagger {\bf S} = {\bf 1}$ and ${\bf V}^\dagger {\bf R} + {\bf S}^\dagger {\bf U} = {\bf 0} $, the diagonal matrices $\widehat{\bf m} = {\rm diag}\left\{m_1^{}, m_2^{}, m_3^{}\right\}$ and $\widehat{\bf M} = {\rm diag}\left\{M_1^{}, M_2^{}, M_3^{}\right\}$ with $m_i^{}$ and $M_i^{}$ (for $i=1,2,3$) being the masses of light and heavy Majorana neutrinos, respectively. At this time, we arrive at the {\it exact} seesaw relation in the mass basis of six Majorana neutrinos:
	\begin{eqnarray}
		\label{eq:SSrelation}
		{\bf V} \widehat{\bf m} {\bf V}^{\rm T} + {\bf R} \widehat{\bf M} {\bf R}^{\rm T} = {\bf 0} \;.
	\end{eqnarray}
	These twelve equations involve relations among all eighteen model parameters. In particular, ${\bf V}$ is the Pontecorvo-Maki-Nakagawa-Sakata (PMNS) matrix~\cite{Pontecorvo:1957cp,Maki:1962mu}, and ${\bf R}$ signifies the strengths of charged-current interactions of heavy Majorana neutrinos.
	
	In fact, diagonalizing a general $6 \times 6$ mass matrix and obtaining the explicit expressions for ${\bf V}$, ${\bf R}$, ${\bf S}$ and ${\bf U}$ are practically challenging. Therefore, a specific parametrization of ${\cal U}$ is necessary to identify physical parameters and to facilitate calculations. For example, an Euler-like block parametrization~\cite{Xing:2007zj,Xing:2011ur,Xing:2020ijf} has been used to build a bridge connecting the {\it original} seesaw parameters with the low-energy {\it derivatives} as well as to discuss the dynamic properties of the seesaw mechanism in a series of work~\cite{Xing:2023adc,Xing:2023kdj,Xing:2024xwb,Xing:2024cyc,Xing:2024gmy,Xing:2025ufi}. Recently, a Wigner-like parametrization has been proposed to express ${\cal U}$ as~\cite{Wigner:1968,Zhou:2025qwk}
	\begin{eqnarray}
		\label{eq:wignerU}
		{\cal U} = \begin{pmatrix}
			{\bf V} & {\bf R} \\ {\bf S} & {\bf U}
		\end{pmatrix} = \begin{pmatrix}
		{\bf u}_1^{} & {\bf 0} \\ {\bf 0} & {\bf u}_2^{} 
		\end{pmatrix} \cdot \begin{pmatrix}
		\widehat{\bf c} & \widehat{\bf s} \\ -\widehat{\bf s} & \widehat{\bf c}
		\end{pmatrix} \cdot \begin{pmatrix}
		{\bf v}_1^{} & {\bf 0} \\ {\bf 0} & {\bf v}_2^{} 
		\end{pmatrix} \;,
	\end{eqnarray}
	where ${\bf u}_1^{}$, ${\bf u}_2^{}$, ${\bf v}_1^{}$ and ${\bf v}_2^{}$ are all $3\times3$ unitary matrices, $\widehat{\bf c} = {\rm diag}\left\{c_1^{}, c_2^{}, c_3^{}\right\}$ and $\widehat{\bf s} = {\rm diag}\left\{s_1^{}, s_2^{}, s_3^{}\right\}$ are diagonal matrices with $c_i^{} \equiv \cos\vartheta_i^{}$ and $s_i^{} \equiv \sin\vartheta_i^{}$ (for $i=1,2,3$) for three rotation angles $\vartheta_i^{} \in [0,\pi/2)$. At this time, the seesaw relation in Eq.~(\ref{eq:SSrelation}) with ${\bf V} = {\bf u}_1^{} \widehat{\bf c} {\bf v}_1^{}$ and ${\bf R} = {\bf u}_1^{} \widehat{\bf s} {\bf v}_2^{}$ becomes
	\begin{eqnarray}
		\label{eq:SSrelationWigner}
		{\bf v}_1^{} \widehat{\bf m} {\bf v}_1^{\rm T} = - \widehat{\bf s} \widehat{\bf c}^{-1} \cdot {\bf v}_2^{} \widehat{\bf M} {\bf v}_2^{\rm T} \cdot \widehat{\bf s} \widehat{\bf c}^{-1} \;.
	\end{eqnarray}
	
	Based on the Lagrangian in Eq.~(\ref{eq:type-I-lag}), it is possible to perform the transformations on the left-handed lepton doublet $\ell_{\rm L}^{} \to {\bf u}_1^{} \ell_{\rm L}^{}$ and on the right-handed neutrino singlet $N_{\rm R}^{} \to {\bf u}_2^* N_{\rm R}^{}$. In this flavor basis, we choose the effective mass matrices $\widetilde{\bf m}_{\rm D}^{} \equiv {\bf u}_1^\dagger {\bf m}_{\rm D}^{} {\bf u}_2^*$ and $\widetilde{\bf m}_{\rm R}^{} \equiv {\bf u}_2^\dagger {\bf m}_{\rm R}^{} {\bf u}_2^*$ to be both diagonal, i.e., $\left[\widetilde{\bf m}_{\rm D}^{}, \widetilde{\bf m}_{\rm R}^{}\right] = {\bf 0}$. Working with the diagonal ${\bf m}_{\rm R}^{}$, the mixing matrix ${\cal U}$ can be expressed in a rather simple way
	\begin{eqnarray}
		\label{eq:U6*6wigner}
		{\cal U} = \begin{pmatrix}
			{\rm i} {\bf u}_1^{} \widehat{\bf c} & {\bf u}_1^{} \widehat{\bf s} \\ - {\rm i} \widehat{\bf s} & \widehat{\bf c}
		\end{pmatrix} \;,
	\end{eqnarray}
	with ${\bf V} = {\rm i} {\bf u}_1^{} \widehat{\bf c}$, ${\bf R} = {\bf u}_1^{} \widehat{\bf s}$, ${\bf u}_2^{} = {\bf 1}$, ${\bf v}_1^{} = {\rm i}{\bf 1}$ and ${\bf v}_2^{} = {\bf 1}$. We can further reconstruct ${\bf m}_{\rm D}^{} = {\bf u}_1^{} \widetilde{\bf m}_{\rm D}^{}$ and ${\bf m}_{\rm R}^{} = \widetilde{\bf m}_{\rm R}^{}$, satisfying $[{\bf m}_{\rm D}^\dagger {\bf m}_{\rm D}^{}, {\bf m}_{\rm R}^\dagger {\bf m}_{\rm R}^{}] = {\bf 0}$. Notice that the imaginary unit in ${\bf v}_1^{}$ and ${\bf V}$ do not appear in observables, since it is unphysical and can be absorbed by redefining three charged-lepton fields.
	
	The effective mass matrices can be parametrized as $\widetilde{\bf m}_{\rm D}^{} = {\rm diag} \left\{D_1^{},D_2^{},D_3^{}\right\}$ and $\widetilde{\bf m}_{\rm R}^{} = {\rm diag}\left\{R_1^{},R_2^{},R_3^{}\right\}$ without loss of generality.\footnote{A similar choice of $\widetilde{\bf m}_{\rm D}^{}$ and $\widetilde{\bf m}_{\rm R}^{}$ is also mentioned in Ref.~\cite{Cheng:1980tp}, in which all three eigenvalues of $\widetilde{\bf m}_{\rm R}^{}$ are imposed to be identical.} Then we have $D_i^{} = m_i^{} / t_i^{}$ and $M_i^{} = m_i^{} / t_i^2$ with $t_i^{} \equiv \tan\vartheta_i^{}$ (for $i=1,2,3$) from Eq.~(\ref{eq:SSrelationWigner}). Since we prefer $D_i^{} \approx {\cal O} \left(\Lambda_{\rm EW}^{}\right)$ and $M_i^{} \approx {\cal O} \left(\Lambda_{\rm SS}^{}\right)$ in seesaw mechanism, the rotation angles satisfy $t_i^{} \approx \vartheta_i^{} \approx {\cal O} \left(\Lambda_{\rm EW}^{} / \Lambda_{\rm SS}^{}\right)$, reflecting the vast hierarchy between two energy scales. Therefore, series expansions with respect to $\vartheta_i^{}$ are always safe and reasonable. For example, one can expand $R_i^{}$ as $R_i^{} = M_i^{} \left(1-t_i^2\right) \approx M_i^{} \left(1-\vartheta_i^2\right)$. This explicitly clarifies that $R_i^{} \neq M_i^{}$ and then ${\bf m}_{\rm R}^{} \neq \widehat{\bf M}$, while the relation $R_i^{} = M_i^{}$ only satisfies at the leading order. In fact, the masses of heavy Majorana neutrinos $M_i^{}$ are actually acquired by diagonalizing the $6\times 6$ mass matrix in Eq.~(\ref{eq:diagonalize}), which are three heavy eigenvalues out of the six. In contrast, $R_i^{}$ are only the eigenvalues of the $3\times 3$ mass matrix ${\bf m}_{\rm R}^{}$ and are not the physical masses of the corresponding heavy Majorana neutrinos in the mass basis {\it after} the SSB.\footnote{Before the SSB, there is no mixing between neutrinos in singlets and in lepton doublets, so the physical masses of three heavy neutrinos are indeed the eigenvalues of ${\bf m}_{\rm R}^{}$. More discussions on this mismatch can be found in Ref.~\cite{Xing:2023adc}.}
	
	The rotation angles $\vartheta_i^{}$ also describe the non-unitarity effects of the leptonic flavor mixing. The PMNS matrix ${\bf V}$ is approximately an unitary matrix at the leading order with ${\bf V} \simeq {\bf u}_1^{}$, which can be parametrized in the standard way advocated by Particle Data Group as~\cite{ParticleDataGroup:2024cfk}
	\begin{eqnarray}
		\label{eq:u1}
		{\bf u}_1^{} = \begin{pmatrix}
			c_{12}^{} c_{13}^{} & s_{12}^{} c_{13}^{} & s_{13}^{} {\rm e}^{-{\rm i} \delta}  \\
			-s_{12}^{} c_{23}^{} - c_{12}^{} s_{13}^{} s_{23}^{} {\rm e}^{{\rm i}\delta} & c_{12}^{} c_{23}^{} - s_{12}^{} s_{13}^{} s_{23}^{} {\rm e}^{{\rm i} \delta} & c_{13}^{} s_{23}^{} \\
			s_{12}^{} s_{23}^{} - c_{12}^{} s_{13}^{} c_{23}^{} {\rm e}^{{\rm i}\delta} & -c_{12}^{} s_{23}^{} - s_{12}^{} s_{13}^{} c_{23}^{} {\rm e}^{{\rm i} \delta} & c_{13}^{} c_{23}^{} 
		\end{pmatrix} \cdot \begin{pmatrix}
			{\rm e}^{{\rm i} \rho} & 0 & 0  \\
			0 & {\rm e}^{{\rm i} \sigma} & 0 \\
			0 & 0 & 1 \end{pmatrix} \;,
	\end{eqnarray}
	with $s_{ij}^{} \equiv \sin\theta_{ij}^{}$ and $c_{ij}^{} \equiv \cos\theta_{ij}^{}$ (for $ij=12,13,23$), $\delta$ being the Dirac CP-violating phase and $\{\rho,\sigma\}$ being two Majorana phases. The unique Jarlskog invariant ${\cal J}$ for the unitary mixing matrix is defined as~\cite{Jarlskog:1985ht,Wu:1985ea,Cheng:1986in}
	\begin{eqnarray}
		{\cal J} \sum_\gamma \epsilon_{\alpha \beta \gamma}^{} \sum_k \epsilon_{ijk}^{} \equiv {\rm Im} \left[\left({\bf u}_1^{}\right)_{\alpha i}^{} \left({\bf u}_1^{}\right)_{\beta j}^{} \left({\bf u}_1^*\right)_{\alpha j}^{} \left({\bf u}_1^*\right)_{\beta i}^{}  \right] \;, 
	\end{eqnarray}
	with ${\cal J} = s_{12}^{} c_{12}^{} s_{13}^{} c_{13}^2 s_{23}^{} c_{23}^{} \sin\delta$ in the standard parametrization of ${\bf u}_1^{}$. Similarly, we can define the Jarlskog-like rephasing invariant $J_{\alpha\beta}^{ij} \equiv {\rm Im} \left({\bf V}_{\alpha i}^{} {\bf V}_{\beta j}^{} {\bf V}_{\alpha j}^{*} {\bf V}_{\beta i}^{*} \right)$ for $\alpha,\beta=e,\mu,\tau$ and $i,j=1,2,3$, satisfying $J_{\alpha\beta}^{ij} = - J_{\beta\alpha}^{ij} = - J_{\alpha\beta}^{ji} = J_{\beta\alpha}^{ji}$ and $J_{\alpha\alpha}^{ij} = J_{\alpha\beta}^{ii} = 0$~\cite{Wang:2021rsi}. The explicit relation between ${\cal J}$ and $J_{\alpha\beta}^{ij}$ takes the form as
	\begin{eqnarray}
		\label{eq:JcalJ}
		J_{\alpha\beta}^{ij} = c_i^2 c_j^2 {\cal J} \sum_\gamma \epsilon_{\alpha \beta \gamma}^{} \sum_k \epsilon_{ijk}^{} \;,
	\end{eqnarray}
	suggesting that all invariants with the same $ij$-index combination are equal and proportional to $c_i^2 c_j^2 {\cal J}$ up to a minus sign. On the other hand, the neutrino-antineutrino oscillation, which ought to happen for Majorana neutrinos, involves another type of invariant~\cite{Xing:2013ty,Xing:2013woa}
	\begin{eqnarray}
		\label{eq:calV}
		{\cal V}_{\alpha\beta}^{ij} \equiv {\rm Im}\left({\bf V}_{\alpha i}^{} {\bf V}_{\beta i}^{} {\bf V}_{\alpha j}^{*} {\bf V}_{\beta j}^{*}\right) = c_i^2 c_j^2 \widetilde{\cal V}_{\alpha\beta}^{ij} \;,
	\end{eqnarray}
	where $\widetilde{\cal V}_{\alpha\beta}^{ij} \equiv {\rm Im}[\left({\bf u}^{}_1\right)_{\alpha i}^{} \left({\bf u}^{}_1\right)_{\beta i}^{} \left({\bf u}^{*}_1\right)_{\alpha j}^{} \left({\bf u}^{*}_1\right)_{\beta j}^{}]$ refers to the invariant in the unitary case, satisfying $\widetilde{\cal V}_{\alpha\beta}^{ij} = \widetilde{\cal V}_{\beta\alpha}^{ij} = - \widetilde{\cal V}_{\alpha\beta}^{ji} = - \widetilde{\cal V}_{\beta\alpha}^{ji}$. Only nine of $\widetilde{\cal V}_{\alpha\beta}^{ij}$ are independent due to the unitarity of ${\bf u}_1^{}$, which can be chosen as $\widetilde{\cal V}_{\alpha\alpha}^{ij}$ for $ij=12,13,23$ and $\alpha=e,\mu,\tau$. Their specific expressions are given in Appendix~\ref{app:A} for reference. As shown in Eqs.~(\ref{eq:JcalJ}) and (\ref{eq:calV}), it yields simple and straightforward expressions for two types of Jarlskog-like invariants with the non-unitary mixing matrix in our case. These invariants are all proportional to their counterparts in the unitary case, while $\vartheta_i^{}$ quantify the derivation from the exact unitarity of the mixing matrix. 
	
	From such a demonstration, we believe that rigorous calculations are feasible within the canonical seesaw model, and all such observables can be expressed by the following selected set of physical parameters: masses of three light Majorana neutrinos $\{m_1^{}, m_2^{}, m_3^{}\}$, three mixing angles $\{\theta_{12}^{}, \theta_{13}^{}, \theta_{23}^{}\}$ and three CP-violating phases $\{\delta, \rho, \sigma\}$ in ${\bf u}_1^{}$, and three introduced rotation angles $\left\{\vartheta_1^{}, \vartheta_2^{}, \vartheta_3^{} \right\}$. Compared to eighteen independent parameters in the full theory, six parameters are missing here. This comes from the fact that the $3\times 3$ Hermitian matrix ${\bf m}_{\rm D}^\dagger {\bf m}_{\rm D}^{}$ is diagonal in our work, and therefore providing six real constraints from three zero off-diagonal elements.

	\section{LFV Decays of Charged Leptons}
	
	\label{sec:mutoegamma}
	
	\begin{figure}[t]
		\centering
		\includegraphics[scale=0.59]{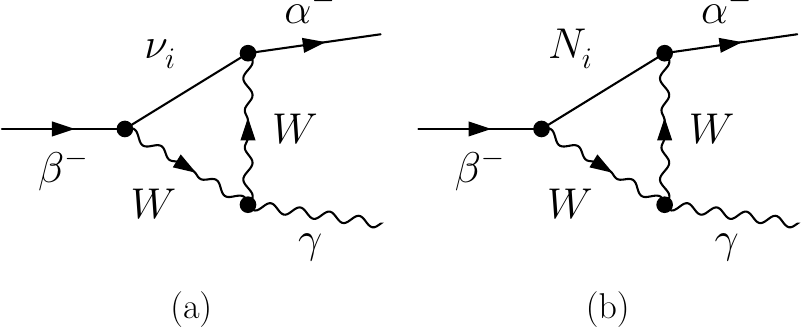}
		\vspace{-0.2cm}
		\caption{The Feynman diagrams of the LFV decay $\beta^- \to \alpha^- + \gamma$ in the canonical seesaw model, mediated by (a) light neutrinos $\nu_i^{}$ and (b) heavy neutrinos $N_i^{}$ for $i=1,2,3$, respectively.}
		\label{fig:LFV}
	\end{figure}
	
	In the canonical seesaw model, charged leptons can undergo LFV radiative decays $\beta^- \to \alpha^- + \gamma$ for $(\alpha,\beta) = (e,\mu), (e,\tau)$ and $(\mu,\tau)$, mediated by both light and heavy Majorana neutrinos as illustrated in Fig.~\ref{fig:LFV}. The rates of such LFV decays read~\cite{Cheng:1980tp,Ilakovac:1994kj,Alonso:2012ji,Xing:2020ivm}
	\begin{eqnarray}
		\Gamma \left(\beta^- \to \alpha^- + \gamma\right) &=& \frac{\alpha_{\rm em}^{} G_\mu^2 m_\beta^5}{128 \pi^4} \left(1+\frac{m_\alpha^2}{m_\beta^2}\right) \left(1-\frac{m_\alpha^2}{m_\beta^2}\right)^3 \nonumber \\
		&& \times \left|\sum_{i=1}^{3} {\bf V}_{\alpha i}^{} {\bf V}_{\beta i}^{*} G_\gamma^{} \left(\frac{m_i^2}{m_W^2} \right) + \sum_{i=1}^{3} {\bf R}_{\alpha i}^{} {\bf R}_{\beta i}^{*} G_\gamma^{} \left(\frac{M_i^2}{m_W^2} \right)\right|^2 \;,
	\end{eqnarray}
	where $\alpha_{\rm em}^{}$ is the electromagnetic fine-structure constant, $G_\mu^{}$ is the Fermi coupling constant, $\{m_\alpha^{}, m_\beta^{}, m_W^{}\}$ are masses of the charged leptons $\alpha^-$, $\beta^-$ and the $W$ boson, respectively. The loop function is defined as
	\begin{eqnarray}
		G_\gamma^{} (x) = - \frac{x(2x^2+5x-1)}{4(1-x)^3} - \frac{3x^3}{2(1-x)^4}\ln x \;.
	\end{eqnarray}
	In our case, as the expressions of ${\bf V}$ and ${\bf R}$ have already been derived, decay rates can be written in a more compact way as
	\begin{eqnarray}
		\Gamma \left(\beta^- \to \alpha^- + \gamma\right) = \frac{\alpha_{\rm em}^{} G_\mu^2 m_\beta^5}{128 \pi^4} \left(1+\frac{m_\alpha^2}{m_\beta^2}\right) \left(1-\frac{m_\alpha^2}{m_\beta^2}\right)^3 \left| \left({\bf u}_1^{} {\bf G}_\gamma^{} {\bf u}_1^\dagger\right)_{\alpha \beta}^{} \right|^2 \;,
	\end{eqnarray}
	where $\widehat{\bf G}_\gamma^{} \equiv {\rm diag}\{{\cal G}_1^{},{\cal G}_2^{},{\cal G}_3^{}\}$ with 
	\begin{eqnarray}
		{\cal G}_i^{} \equiv c_i^2 G_\gamma^{} \left(\frac{m_i^2}{m_W^2}\right) + s_i^{2} G_\gamma^{} \left(\frac{M_i^{2}}{m_W^2}\right) \;,
	\end{eqnarray}
	and heavy Majorana neutrino masses being $M_i^{} = m_i^{} / t_i^2$ (for $i=1,2,3$). In the standard parametrization of ${\bf u}_1^{}$, the decay rate of $\mu^- \to e^- + \gamma$ is
	\begin{eqnarray}
		\label{eq:muegamma}
		\Gamma \left(\mu^- \to e^- + \gamma\right) &=& \frac{\alpha_{\rm em}^{} G_\mu^2 m_\mu^5}{128 \pi^4} \left(1+\frac{m_e^2}{m_\mu^2}\right) \left(1-\frac{m_e^2}{m_\mu^2}\right)^3 \nonumber \\
		&& \times \left\{ {\cal G}_1^2 \left(s^2_{12} c^2_{12} c^2_{13} c^2_{23} + 2 s_{12}^{} c^3_{12} s_{13}^{} c^2_{13} s_{23}^{} c_{23}^{} c_\delta^{} + c^4_{12} s^2_{13} c^2_{13} s^2_{23} \right) \right. \nonumber \\
		&& + {\cal G}_2^2 \left(s^2_{12} c^2_{12} c^2_{13} c^2_{23} -2 s^3_{12} c_{12}^{} s_{13}^{} c^2_{13} s_{23}^{} c_{23}^{} c_\delta^{} + s^4_{12} s^2_{13} c^2_{13} s^2_{23} \right) + {\cal G}_3^2 \left(s^2_{13} c^2_{13} s^2_{23}\right) \nonumber \\
		&& + 2 {\cal G}_1^{} {\cal G}_2^{} \left[s^{}_{12} c_{12}^{} s_{13}^{} c^2_{13} s_{23}^{} c_{23}^{} c_\delta^{} \left(s_{12}^2 - c_{12}^2\right) - s^2_{12} c^2_{12} c^2_{13} \left(c^2_{23} - s^2_{13} s^2_{23} \right) \right] \nonumber \\
		&& - 2 {\cal G}_1^{} {\cal G}_3^{} \left(s_{12}^{} c_{12}^{} s_{13}^{} c^2_{13} s_{23}^{} c_{23}^{} c_\delta^{} + c^2_{12} s^2_{13} c^2_{13} s^2_{23} \right) \nonumber \\
		&& \left. + 2 {\cal G}_2^{} {\cal G}_3^{} \left(s_{12}^{} c_{12}^{} s_{13}^{} c^2_{13} s_{23}^{} c_{23}^{} c_\delta^{} - s^2_{12} s^2_{13} c^2_{13} s^2_{23} \right)  \right\} \;,
	\end{eqnarray}
	with $s_\delta^{} \equiv \sin\delta$ and $c_\delta^{} \equiv \cos\delta$. The rate for the tau decay $\tau^- \to e^- + \gamma$ can be similarly obtained by replacing $m_\mu^{} \to m_\tau^{}$, $s_{23}^{} \leftrightarrow c_{23}^{}$ and $c_{\delta}^{} \to - c_{\delta}^{}$ in Eq.~(\ref{eq:muegamma}) on account of the $\mu$-$\tau$ permutation symmetry~\cite{Xing:2015fdg,Xing:2022uax}. Finally, the remaining decay channel $\tau^- \to \mu^- + \gamma$ owns the decay rate as 
	\begin{eqnarray}
		\Gamma \left(\tau^- \to \mu^- + \gamma\right) &=& \frac{\alpha_{\rm em}^{} G_\mu^2 m_\tau^5}{128 \pi^4} \left(1+\frac{m_\mu^2}{m_\tau^2}\right) \left(1-\frac{m_\mu^2}{m_\tau^2}\right)^3 \nonumber \\
		&& \times \left\{ {\cal G}_1^2 \left\{2 s_{12}^{} c_{12}^{} s_{13}^{} s^2_{23} c_{23}^{} c_{2\delta}^{} \left[s_{23}^{} c_\delta^{} \left(s^2_{12} - c^2_{12} s^2_{13} \right) - s_{12}^{} c_{12}^{} s_{13}^{} c_{23}^{} \right] \right. \right. \nonumber \\
		&& + 2 s_{12}^{} c_{12}^{} s_{13}^{} s_{23}^{} c_{23}^{} c_\delta^{} \left(c^2_{23} - 2 s^2_{23} s^2_\delta \right) \left(c^2_{12} s^2_{13} - s^2_{12} \right) + s^2_{12} c^2_{12} s^2_{13} \left(c^2_{23} - s^2_{23} \right)^2 \nonumber \\
		&& \left.  + s^2_{23} c^2_{23} \left(c^4_{12} s^4_{13} + s^4_{12} \right) \right\} \nonumber \\
		&& + {\cal G}_2^2 \left\{-2  s^{}_{12} c^{}_{12} s^{}_{13} s^2_{23} c^{}_{23} c_{2 \delta}^{} \left[s^{}_{23} c_\delta^{} \left(c^2_{12} - s^2_{12} s^2_{13} \right) + s^{}_{12} c^{}_{12} s^{}_{13} c^{}_{23} \right] \right. \nonumber \\
		&& + s^2_{12} c^2_{12} s^2_{13} \left(c^2_{23} - s^2_{23} \right)^2 - 2 s^{}_{12} c^{}_{12} s^{}_{13} s^3_{23} c^{}_{23} s^{}_\delta s^{}_{2\delta} \left(c^2_{12}-s^2_{12} s^2_{13}\right) \nonumber \\
		&& \left. + 2 s^{}_{12} c^{}_{12} s^{}_{13} s^{}_{23} c^3_{23} c^{}_\delta \left(c^2_{12} - s^2_{12} s^2_{13} \right) + s^2_{23} c^2_{23} \left(s^4_{12} s^4_{13} + c^4_{12}\right) \right\} + {\cal G}_3^2 \left(c^4_{13} s^2_{23} c^2_{23} \right) \nonumber \\
		&& - 2 {\cal G}_1^{} {\cal G}_2^{} \left\{s_{13}^2 s_{23}^2 c_{23}^2 \left(s_{12}^4 + c_{12}^4\right) + s_{12}^2 c_{12}^2 \left[s_{13}^2 c_{23}^4 + s_{13}^2 s_{23}^4 - s_{23}^2 c_{23}^2 \left(s_{13}^4 + 1 \right) \right] \right. \nonumber \\
		&& - s_{12}^{} c_{12}^{} s^{}_{13} s^2_{23} c^{}_{23} c^{}_{2\delta} \left[2 s^{}_{12} c^{}_{12} s^{}_{13} c^{}_{23} c^2_\delta + 2 s^{}_{12} c^{}_{12} s^{}_{13} c^{}_{23} s^2_\delta \right. \nonumber \\
		&& \left. + s^{}_{23} c_\delta^{} \left(s^2_{12} + 1 \right)  \left(c^2_{12} - s^2_{12} \right) \right] - 2  s^{}_{12} c^{}_{12} s^{}_{13} s^3_{23} c^{}_{23} s^2_\delta c_{\delta}^{} \left(s^2_{13} + 1 \right) \left(c^2_{12} - s^2_{12} \right) \nonumber \\
		&& \left. + 2 s_{12}^{} c^{}_{12} s^{}_{13} s^{}_{23} c^3_{23} c_\delta^{} \left(s^2_{13} + 1 \right)  \left(s^2_{12} - c^2_{12} \right) \right\} \nonumber \\
		&& - 2 {\cal G}_1^{} {\cal G}_3^{} \left[ s^{}_{12} c^{}_{12} s^{}_{13} c^2_{13} s^3_{23} c^{}_{23} c_\delta^{} c^{}_{2\delta} + s^{}_{12} c^{}_{12} s^{}_{13} c^2_{13} s^3_{23} c^{}_{23} s^{}_\delta s^{}_{2\delta} \right. \nonumber \\
		&& \left. - s^{}_{12} c^{}_{12} s^{}_{13} c^2_{13} s^{}_{23} c^3_{23} c^{}_{\delta} + c^2_{13} s^2_{23} c^2_{23} \left(s^2_{12} - c^2_{12} s^2_{13} \right) \right] \nonumber \\
		&& + 2 {\cal G}_2^{} {\cal G}_3^{} \left( s^{}_{12} c^{}_{12} s^{}_{13} c^2_{13} s^3_{23} c^{}_{23} c^{}_{\delta} c^{}_{2\delta} +  s^{}_{12} c^{}_{12} s^{}_{13} c^2_{13} s^3_{23} c^{}_{23} s^{}_\delta s^{}_{2\delta} \right. \nonumber \\
		&& \left.\left. - s^{}_{12} c^{}_{12} s^{}_{13} c^2_{13} s^{}_{23} c^3_{23} c^{}_\delta - c^2_{12} c^2_{13} s^2_{23} c^2_{23} + s^2_{12} s^2_{13} c^2_{13} s^2_{23} c^2_{23} \right)  \right\} \;,
	\end{eqnarray}
	with $s_{2\delta}^{} \equiv \sin 2\delta$ and $c_{2\delta}^{} \equiv \cos 2\delta$. Two Majorana phases do not appear in these processes which conserve the total lepton number. All the information is contained in $\vartheta_i^{}$, except for terms related to the mixing angles and the CP-violating phase.
	
	From the experimental aspect, limitations are imposed on the branching fractions ${\cal B}(\beta^- \to \alpha^- + \gamma) \equiv \Gamma(\beta^- \to \alpha^- + \gamma) / \Gamma_{\rm total} $. The LFV decay rates against those of the purely leptonic decays $\beta^- \to \alpha^- + \overline{\nu}_\alpha^{} + \nu_\beta^{}$ are described by the dimensionless parameters
	\begin{eqnarray}
		\xi_{\alpha\beta}^{} \equiv \frac{\Gamma \left(\beta^- \to \alpha^- + \gamma\right)}{\Gamma \left(\beta^- \to \alpha^- + \overline{\nu}_\alpha^{} + \nu_\beta^{} \right)} = \frac{{\cal B} \left(\beta^- \to \alpha^- + \gamma\right)}{{\cal B} \left(\beta^- \to \alpha^- + \overline{\nu}_\alpha^{} + \nu_\beta^{} \right)} \;.
	\end{eqnarray}
	In the non-unitary case, the decay rates of $\beta^- \to \alpha^- + \overline{\nu}_\alpha^{} + \nu_\beta^{}$ are given by
	\begin{eqnarray}
		\Gamma \left(\beta^- \to \alpha^- + \overline{\nu}_\alpha^{} + \nu_\beta^{} \right) = \frac{G_{\mu}^2 m_\beta^5}{192 \pi^3}\left(1-8 \frac{m_\alpha^2}{m_\beta^2}\right)\left[1+\frac{\alpha^{}_{\rm em}}{2 \pi}\left(\frac{25}{4}-\pi^2\right)\right] \sum_{i,j=1}^3 \left|{\bf V}_{\alpha i}^{}\right|^2 \left|{\bf V}_{\beta j}^{}\right|^2 \;,
	\end{eqnarray}
	where we have included the QED corrections at the one-loop level. Then we arrive at
	\begin{eqnarray}
		\label{eq:muenunu}
		\Gamma \left(\mu^- \to e^- + \overline{\nu}_e^{} + \nu_\mu^{} \right) &=& \frac{G_{\mu}^2 m_\mu^5}{192 \pi^3}\left(1-8 \frac{m_e^2}{m_\mu^2}\right)\left[1+\frac{\alpha^{}_{\rm em}}{2 \pi}\left(\frac{25}{4}-\pi^2\right)\right] \nonumber \\
		&& \times \left[c_1^4 c^2_{12} c^2_{13} \left(c^2_{12} s^2_{13} s^2_{23} + 2 s^{}_{12} c^{}_{12} s^{}_{13} s^{}_{23} c^{}_{23} c^{}_\delta  + s^2_{12} c^2_{23}\right) \right. \nonumber \\
		&& + c_2^4 s^2_{12} c^2_{13} \left(s^2_{12} s^2_{13} s^2_{23} - 2 s^{}_{12} c^{}_{12} s^{}_{13} s^{}_{23} c^{}_{23} c_\delta^{} + c^2_{12} c^2_{23}\right) + c_3^4 s^2_{13} c^2_{13} s^2_{23} \nonumber \\
		&& + c_1^2 c_2^2  c^2_{13} \left(2 s^3_{12} c^{}_{12} s^{}_{13} s^{}_{23} c_{23}^{} c^{}_\delta - 2 s^{}_{12} c^3_{12} s^{}_{13} s^{}_{23} c^{}_{23} c^{}_\delta + 2 s^2_{12} c^2_{12} s^2_{13} s^2_{23} \right. \nonumber \\
		&& \left. + c^4_{12} c^2_{23}+s^4_{12} c^2_{23}\right)  + c_1^2 c_3^2 \left(s^2_{12} s^2_{13} c^2_{23} + 2s^{}_{12} c^{}_{12} s^3_{13} s^{}_{23} c^{}_{23} c_\delta^{}  \right. \nonumber \\
		&& \left. + c^2_{12} s^4_{13} s^2_{23} +c^2_{12} c^4_{13} s^2_{23}\right) + c_2^2 c_3^2 \left(s^2_{12} s^4_{13} s^2_{23} -2 s^{}_{12} c^{}_{12} s^3_{13} s^{}_{23} c^{}_{23} c_\delta^{} \right. \nonumber \\
		&& \left. \left. +s^2_{12} c^4_{13} s^2_{23}+c^2_{12} s^2_{13} c^2_{23}\right)\right] \;.
	\end{eqnarray}
	The decay rate for $\tau^- \to e^- + \overline{\nu}_e^{} + \nu_\tau^{}$ can be directly written down from Eq.~(\ref{eq:muenunu}) by changing $m_\mu^{} \to m_\tau^{}$, $s_{23}^{} \leftrightarrow c_{23}^{}$ and $c_{\delta}^{} \to - c_{\delta}^{}$ as before. For $\tau^- \to \mu^- + \overline{\nu}_\mu^{} + \nu_\tau^{}$ we have 
	\begin{eqnarray}
		\Gamma \left(\tau^- \to \mu^- + \overline{\nu}_\mu^{} + \nu_\tau^{} \right) &=& \frac{G_{\mu}^2 m_\tau^5}{192 \pi^3}\left(1-8 \frac{m_\mu^2}{m_\tau^2}\right)\left[1+\frac{\alpha^{}_{\rm em}}{2 \pi}\left(\frac{25}{4}-\pi^2\right)\right] \nonumber \\
		&& \times \left\{c_1^4 c^2_{12} \left(c^2_{12} s^2_{13} c^2_{23}-2 s^{}_{12} c^{}_{12} s^{}_{13} s^{}_{23} c^{}_{23} c^{}_\delta + s^2_{12} s^2_{23}\right) \right. \nonumber \\
		&& \times \left(c^2_{12} s^2_{13} s^2_{23} + 2  s^{}_{12} c^{}_{12} s^{}_{13} s^{}_{23} c^{}_{23} c^{}_\delta + s^2_{12} c^2_{23}\right) \nonumber \\
		&& + c_2^4 \left(s^2_{12} s^2_{13} c^2_{23} + 2 s^{}_{12} c^{}_{12} s^{}_{13} s^{}_{23} c^{}_{23} c^{}_\delta + c^2_{12} s^2_{23}\right) \nonumber \\
		&& \times \left(s^2_{12} s^2_{13} s^2_{23} - 2 s^{}_{12} c^{}_{12} s^{}_{13} s^{}_{23} c^{}_{23}c^{}_\delta + c^2_{12} c^2_{23}\right) + c_3^4 c^4_{13} s^2_{23} c^2_{23} \nonumber \\
		&& + c_1^2 c_2^2 \left[\left(s^2_{12} s^2_{13} c^2_{23} + 2 s^{}_{12} c^{}_{12} s^{}_{13} s^{}_{23} c^{}_{23} c^{}_\delta + c^2_{12} s^2_{23}\right) \right. \nonumber \\
		&& \times \left(c^2_{12} s^2_{13} s^2_{23} + 2 s^{}_{12} c^{}_{12} s^{}_{13} s^{}_{23} c^{}_{23} c^{}_\delta + s^2_{12} c^2_{23}\right) \nonumber \\
		&& + \left(c^2_{12} s^2_{13} c^2_{23} - 2 s^{}_{12} c^{}_{12} s^{}_{13} s^{}_{23} c^{}_{23} c^{}_\delta + s^2_{12} s^2_{23}\right) \nonumber \\
		&& \left. \times \left(s^2_{12} s^2_{13} s^2_{23}-2 s^{}_{12} c^{}_{12} s^{}_{13} s^{}_{23} c^{}_{23}c^{}_\delta + c^2_{12} c^2_{23}\right)\right] \nonumber \\
		&& + c_1^2 c_3^2 c^2_{13} \left(2 c^2_{12} s^2_{13} s^2_{23} c^2_{23} + 2 s^{}_{12} c^{}_{12} s^{}_{13} s^{}_{23} c^3_{23} c^{}_\delta - 2 s^{}_{12} c^{}_{12} s^{}_{13} s^3_{23} c^{}_{23} c^{}_\delta  \right. \nonumber \\
		&& \left. + s^2_{12} s^4_{23}+s^2_{12} c^4_{23}\right) + c_2^2 c_3^2 \left(2 s^2_{12} s^2_{13} c^2_{13} s^2_{23} c^2_{23} + 2 s^{}_{12} c^{}_{12} s^{}_{13} c^2_{13} s^3_{23} c^{}_{23} c^{}_\delta \right. \nonumber \\
		&& \left.\left.  - 2  s^{}_{12} c^{}_{12} s^{}_{13} c^2_{13} s^{}_{23} c^3_{23} c^{}_\delta + c^2_{12} c^2_{13} c^4_{23} + c^2_{12} c^2_{13} s^4_{23}\right) \right\} .
	\end{eqnarray}
	The above expressions, although seemingly lengthy, are {\it exact} results expressed by the physical parameters without any approximation. Even so, to clarify the dependence behavior of unknown parameters, expanding in terms of $\vartheta_i^{}$ remains valid, where the dimensionless ratio reads
	\begin{eqnarray}
		\xi_{\alpha\beta}^{} \simeq \frac{3 \alpha^{}_{\rm em}}{2 \pi} \sum_{i=1}^3 \left|\left({\bf u}_1^{}\right)_{\alpha i}^{} \left({\bf u}_1^{*}\right)_{\beta i}^{}\right|^2 \left[\frac{1}{4}\left(\frac{m_i^2}{m_W^2}\right) + \frac{\vartheta_i^2}{2} - \frac{\vartheta_i^2}{4} \left(\frac{m_i^2}{m_W^2}\right) \right]^2 \;.
	\end{eqnarray}
	We have neglected all small mass ratios due to the strong hierarchy of charged-lepton masses $m_e^{} \ll m_\mu^{} \ll m_\tau^{}$, and radiative corrections at the next-to-leading order. For the loop function, $G_\gamma^{}(x) \simeq x/4$ when $x\ll1$ and $G_\gamma^{}(x) \simeq 1/2 $ when $x\gg 1$ are good approximations. We notice that such LFV decays are of order ${\cal O}(\vartheta_i^{4})$ and suppressed by the mass ratios ${\cal O}(m_i^2/M_i^2)$.
	
	The next step is to evaluate $\xi^{}_{\alpha\beta}$ and constrain the corresponding parameter space of $\{\vartheta_1^{},\vartheta_2^{},\vartheta_3^{}\}$ with the current upper bounds on branching fractions of $\beta^- \to \alpha^- + \gamma$ and $\beta^- \to \alpha^- + \overline{\nu}_\alpha^{} + \nu_\beta^{} $~\cite{ParticleDataGroup:2024cfk}. That is,
	\begin{eqnarray}
		{\cal B}\left(\mu^- \to e^- + \gamma\right) &<& 4.2 \times 10^{-13} \;, \nonumber \\
		{\cal B}\left(\tau^- \to e^- + \gamma\right) &<& 3.3 \times 10^{-8} \;, \nonumber \\
		{\cal B}\left(\tau^- \to \mu^- + \gamma\right) &<& 4.2 \times 10^{-8} \;,
	\end{eqnarray}
	for LFV decays at the $90\%$ confidence level, and
	\begin{eqnarray}
		{\cal B}\left(\mu^- \to e^- + \overline{\nu}_e^{} + \nu_\mu^{} \right) &\approx& 100\% \;, \nonumber \\
		{\cal B}\left(\tau^- \to e^- + \overline{\nu}_e^{} + \nu_\tau^{} \right) &\approx& 17.82\% \;, \nonumber \\
		{\cal B}\left(\tau^- \to \mu^- + \overline{\nu}_\mu^{} + \nu_\tau^{} \right) &\approx& 17.39\% \;,
	\end{eqnarray}
	for the pure leptonic decays. Through the definition, the values of the ratio $\xi_{\alpha\beta}^{}$ satisfy
	\begin{eqnarray}
		\label{eq:xi_limit}
		\xi_{e\mu}^{} < 4.2 \times 10^{-13} \;, \quad \xi_{e\tau}^{} < 1.85 \times 10^{-7} \;, \quad \xi_{\mu\tau}^{} < 2.42 \times 10^{-7} \;.
	\end{eqnarray}
	The numerical values of other physical parameters are given below:
	
	\begin{figure}[t]
		\centering
		\includegraphics[scale=0.6]{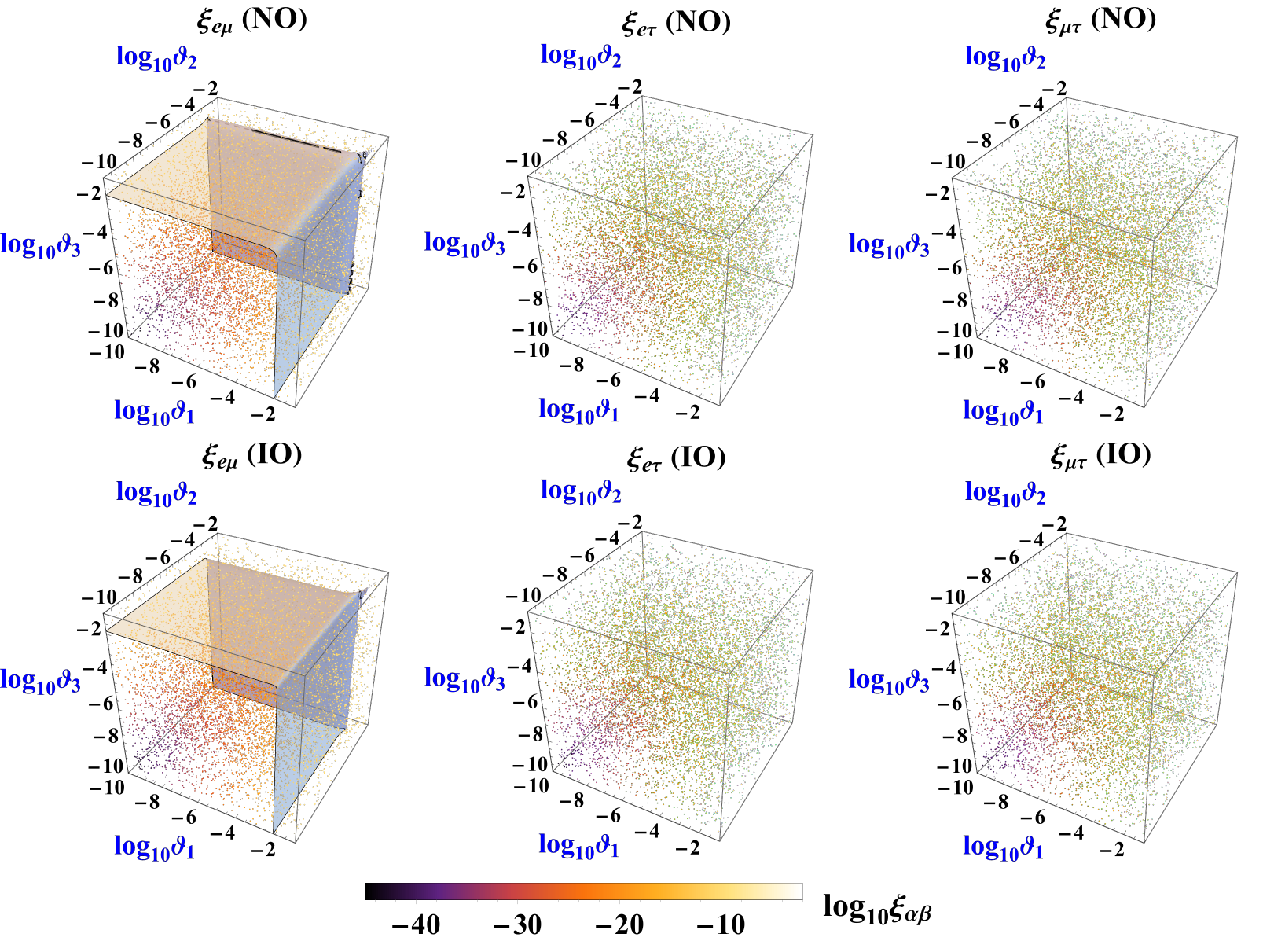}
		\vspace{-0.2cm}
		\caption{The values of $\xi_{\alpha\beta}^{}$ in the case of NO (upper row) and IO (lower row) as the function of $\vartheta_i^{} \in [10^{-10},10^{-1}]$ (for $i=1,2,3$). The left, middle and right columns correspond to cases of $(\alpha,\beta) = (e,\mu), (e,\tau)$ and $(\mu,\tau)$, respectively. Darker colors indicate smaller values of $\xi_{\alpha\beta}^{}$. The light-blue contour surfaces represent the experimental upper bounds listed in Eq.~(\ref{eq:xi_limit}).}
		\label{fig:xi}
	\end{figure}
	
	\begin{itemize}
		\item The electromagnetic fine-structure constant $\alpha^{}_{\rm em} = 1/137.035999084$ and the $W$-boson mass $m_W^{} = 80.369~{\rm GeV}$~\cite{ParticleDataGroup:2024cfk};
		
		\item Three leptonic flavor mixing angles and the Dirac CP-violating phase are extracted from the latest global analysis of the neutrino oscillation data~\cite{Esteban:2024eli,Capozzi:2025wyn,deSalas:2020pgw}. We set the lightest neutrino mass as $m_{\rm lightest}^{} = 5~{\rm meV}$ and arrive at
		\begin{eqnarray}
			m_1^{} &=& 5~{\rm meV}\;, \quad m_2^{} \approx 10~{\rm meV} \;, \quad m_3^{} \approx 50.38~{\rm meV} \;, \nonumber \\
			\theta_{12}^{} &=& 33.68^\circ \;, \quad \theta_{13}^{} = 8.56^\circ \;, \quad \theta_{23}^{} = 43.3^\circ \;, \quad \delta = 212^\circ \;,
		\end{eqnarray}
		for the normal mass ordering of three light neutrinos (NO, $m_1^{} < m_2^{} < m_3^{}$), and 
		\begin{eqnarray}
			m_3^{} &=& 5~{\rm meV}\;, \quad m_1^{} \approx 49.37~{\rm meV} \;, \quad m_2^{} \approx 50.09~{\rm meV} \;, \nonumber \\
			\theta_{12}^{} &=& 33.68^\circ \;, \quad \theta_{13}^{} = 8.59^\circ \;, \quad \theta_{23}^{} = 47.9^\circ \;, \quad \delta = 274^\circ \;,
		\end{eqnarray}
		for the inverted mass ordering (IO, $m_3^{} < m_1^{} < m_2^{}$). 
	\end{itemize}
	
	By randomly sampling the parameters $\vartheta_i^{}$ within the range of $10^{-10} \lesssim \vartheta_i^{} \lesssim 10^{-1}$ (for $i=1,2,3$), we evaluate $\xi_{\alpha\beta}^{}$ in Fig.~\ref{fig:xi} for the NO case (upper row) and the IO case (lower row), where darker colors indicate smaller values of $\xi_{\alpha\beta}^{}$. As $\vartheta_i^{}$ decrease, the value of $\xi_{\alpha\beta}^{}$ correspondingly diminish. The light-blue contour surfaces display the upper bounds of $\xi_{\alpha\beta}^{}$ given in Eq.~(\ref{eq:xi_limit}), while the enclosed region is the parameter space satisfying the latest experimental constraints. Among three decay channels, the most strict constraints to $\vartheta_i^{}$ come from the muon decay $\mu^- \to e^- + \gamma$ with
	\begin{eqnarray}
		\vartheta_1^{} \lesssim 8.93 \times 10^{-3} \;, \quad \vartheta_2^{} \lesssim 7.82 \times 10^{-3} \;, \quad \vartheta_3^{} \lesssim 1.47 \times 10^{-2}
	\end{eqnarray}
	in the NO case, and 
	\begin{eqnarray}
		\vartheta_1^{} \lesssim 8.28 \times 10^{-3} \;, \quad \vartheta_2^{} \lesssim 8.47 \times 10^{-3} \;, \quad \vartheta_3^{} \lesssim 1.42 \times 10^{-2}
	\end{eqnarray}
	in the IO case. The upper bounds on $\vartheta_i^{}$ obtained from tau decays are about two orders of magnitude larger, so there is no contour showing in the plot for the chosen regions of $\vartheta_i^{}$.
	
	\section{CP-violating Asymmetries in Resonant Leptogenesis}
	
	\label{sec:leptogenesis}
	
	\begin{figure}[t]
		\centering
		\includegraphics[scale=0.7]{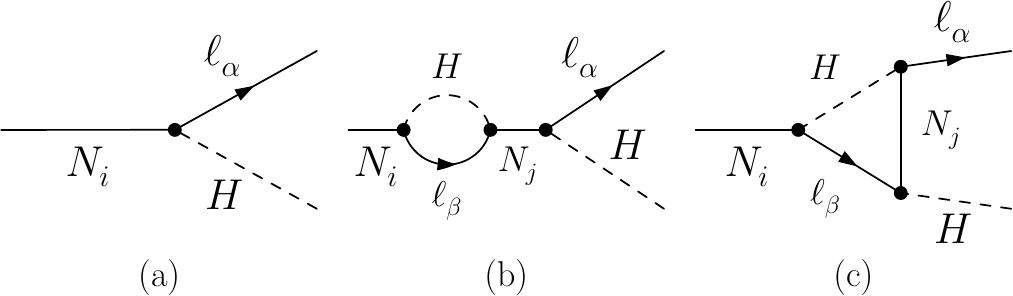}
		\vspace{-0.2cm}
		\caption{Feynman diagrams of the decay $N_i^{} \to \ell_\alpha^{} + H$ at (a) the tree level and (b)-(c) the one-loop level for $i=1,2,3$ and $\alpha,\beta=e,\mu,\tau$.}
		\label{fig:leptogenesis}
	\end{figure}
	
	In thermal leptogenesis, the CP-violating asymmetries between decays $N_i^{} \to \ell_\alpha^{} + H$ and $N_i^{} \to \overline{\ell}_\alpha^{} + \overline{H}$ are generated through the interference between the tree- and loop-level amplitudes (as shown in Fig.~\ref{fig:leptogenesis}), which read~\cite{Luty:1992un,Plumacher:1996kc,Covi:1996wh}:
    \begin{eqnarray}
        \label{eq:epsilon}
        \varepsilon_{i \alpha}^{} &\equiv& \frac{\Gamma\left(N_i^{} \rightarrow \ell_\alpha^{} + H \right) - \Gamma\left(N_i^{} \rightarrow \overline{\ell}_\alpha^{} + \overline{H} \right)}{\sum_\alpha \left[\Gamma\left(N_i^{} \rightarrow \ell_\alpha^{}+H \right)+\Gamma\left(N_i^{} \rightarrow \overline{\ell}_\alpha^{} + \overline{H}\right)\right]} \nonumber \\
        &=& \frac{1}{4 \pi v^2\left({\bf m}_{\rm D}^{\dagger} {\bf m}_{\rm D}^{}\right)_{i i}^{}} \sum_{j \neq i}\left\{{\rm Im}\left[\left({\bf m}_{\rm D}^*\right)_{\alpha i}^{} \left({\bf m}_{\rm D}^{}\right)_{\alpha j}^{} \left({\bf m}_{\rm D}^{\dagger} {\bf m}_{\rm D}^{}\right)_{i j}^{}\right] F\left(\frac{M_j^2}{M_i^2}\right)\right. \nonumber \\
        && \left.+{\rm Im} \left[\left({\bf m}_{\rm D}^*\right)_{\alpha i}^{} \left({\bf m}_{\rm D}^{}\right)_{\alpha j}^{} \left({\bf m}_{\rm D}^{\dagger} {\bf m}_{\rm D}^{} \right)_{i j}^*\right] G\left(\frac{M_j^2}{M_i^2}\right)\right\} \;,
    \end{eqnarray}
    with loop functions
    \begin{eqnarray}
        F(x) \equiv \sqrt{x} \left[1+\frac{1}{1-x}+(1+x)\ln \left(\frac{x}{1+x}\right)\right]\;, \quad G(x) \equiv \frac{1}{1-x} \;.
    \end{eqnarray}
    The above expressions are calculated in the basis where both right-handed neutrino and charged-lepton mass matrices are diagonal, which is also the choice in this work. Generally speaking, the CP violation in the seesaw model is independent of the parametrization. However, the CP violation will vanish when some specific structures are taken for ${\bf m}_{\rm D}^{}$, and this is sometimes irrelevant with the parametrization. For example, on the one hand, the off-diagonal matrix elements of ${\bf m}_{\rm D}^\dagger {\bf m}_{\rm D}^{}$ is zero in our {\it ansatz}, so there is no CP violation for the usual thermal leptogenesis.\footnote{This also corresponds to the case in which the complex orthogonal matrix $O={\bf 1}$ in the Casas-Ibarra parametrization~\cite{Casas:2001sr}.} On the other hand, the CP violation is related to ${\rm Im}\left({\rm det}[{\bf m}_{\rm D}^\dagger {\bf m}_{\rm D}^{}, {\bf m}_{\rm R}^\dagger {\bf m}_{\rm R}^{}]\right)$, so $\varepsilon_{i\alpha}^{} = 0$ when $[{\bf m}_{\rm D}^\dagger {\bf m}_{\rm D}^{}, {\bf m}_{\rm R}^\dagger {\bf m}_{\rm R}^{}] = {\bf 0}$ holds~\cite{Wang:2014lla}. The latter analysis is a general conclusion without any specific parametrization, but requiring to satisfy the commutation relations.
    
    It is still possible to reproduce the CP violation through {\it flavored resonant} thermal leptogenesis by taking the one-loop renormalization-group equations (RGEs) effects into consideration~\cite{Xing:2020erm,Xing:2020ghj,Zhao:2020bzx}, when masses of three heavy Majorana neutrinos satisfy $M_1^{} \simeq M_2^{} \ll M_3^{}$. At this time, the Dirac mass matrix with radiative corrections is given by~\cite{Xing:2020erm}
	\begin{eqnarray}
		\label{eq:CI-RGE}
		{\bf m}_{\rm D}^{} \left(\Lambda_{\rm SS}^{}\right) = {\rm i} \sqrt{I_{\kappa}^{-1}} T_\ell^{-1} {\bf u}_1^{} \left(\Lambda_{\rm EW}^{} \right) \sqrt{\widehat{\bf m}\left(\Lambda_{\rm EW}^{}\right)} \sqrt{\widehat{\bf M}\left(\Lambda_{\rm SS}^{}\right)} \;, 
	\end{eqnarray}
	where $T_\ell^{} \approx {\rm diag}\left\{1,1,1+\Delta_\tau^{}\right\}$ with the strong hierarchy of charged-lepton masses. The evolution functions $\Delta_\tau^{}$ and $I_\kappa^{}$ are defined as
	\begin{eqnarray}
		\Delta_\tau^{} &\equiv& \frac{3}{32\pi^2}\int_{0}^{\ln\left(\Lambda_{\rm SS}^{}/\Lambda_{\rm EW}^{}\right)} y_\tau^2(t)\;{\rm d}t \;, \nonumber \\
		I_\kappa^{} &\equiv& \exp\left\{-\frac{1}{16\pi^2} \int_{0}^{\ln\left(\Lambda_{\rm SS}^{}/\Lambda_{\rm EW}^{}\right)} \left[-3g_2^2(t) + 6 y_t^2(t) + \lambda (t)\right] \;{\rm d}t \right\} \;,
	\end{eqnarray}
	where $t \equiv \ln\left(\mu/\Lambda_{\rm EW}^{}\right)$ with $\mu$ being an arbitrary energy scale between $\Lambda_{\rm EW}^{}$ and $\Lambda_{\rm SS}^{}$, and $\left\{g_2^{},\lambda,y_t^{}\right\}$ being the ${\rm SU}(2)_{\rm L}^{}$ gauge coupling, the Higgs self-coupling and the top-quark Yukawa coupling constant, respectively. Then the CP-violating asymmetries in Eq.~(\ref{eq:epsilon}) turn out to be~\cite{Pilaftsis:1997jf,Pilaftsis:2003gt,Xing:2020ghj}
	\begin{eqnarray}
		\label{eq:epsilon-resonant}
		\varepsilon_{i\alpha}^{} \simeq 2 \Delta_\tau^{} \frac{\left(M_i^2 - M_j^2\right) M_i^{} \Gamma^{}_j}{\left(M_i^2 - M_j^2 \right)^2 + M_i^2 \Gamma_j^2} \times 
		\begin{cases}
			\widetilde{\cal V}_{e\tau}^{ij} + (-1)^{i} {\cal J} M_i^{} / M_j^{}  & \alpha = e \\
			\widetilde{\cal V}_{\mu\tau}^{ij} + (-1)^{i+1} {\cal J} M_i^{} / M_j^{}  & \alpha = \mu \\
			\widetilde{\cal V}_{\tau\tau}^{ij}  & \alpha = \tau \\
		\end{cases} \;,
	\end{eqnarray}
	with $\Gamma_j^{} \simeq I_\kappa^{-1} m_j^{} M_j^2/\left(4\pi v^2\right)$ for $i \neq j = 1,2$. The asymmetries also depend on $\Delta_\tau^{}$ and the Jarlskog-like invariants $\widetilde{\cal V}_{\alpha\tau}^{ij}$ and ${\cal J}$ for the unitary matrix. One may arrive at the expressions of $\widetilde{\cal V}_{e\tau}^{12}$, $\widetilde{\cal V}_{\mu\tau}^{12}$ and $\widetilde{\cal V}_{\tau\tau}^{12}$ with the equations in Appendix~\ref{app:A}, which all depend on $\phi \equiv \rho-\sigma$. Then $\varepsilon_{i\alpha}^{}$ can be {\it exactly} expressed with the input parameters $\{m_i^{}, \vartheta_i^{}, \theta_{ij}^{},\delta,\phi\}$. For a ballpark feeling of these expressions, we expand them with respect to three rotation angles to the leading order, such as
	\begin{eqnarray}
		\varepsilon_{1 e}^{} \simeq \frac{8 \pi v^2 I_\kappa^{} \Delta_\tau^{}}{D_1^{} D_2^4 \vartheta_1^2 \vartheta_2^{}} \left(D_1^2 \vartheta_2^2 - D_2^2 \vartheta_1^2\right) \left(\widetilde{\cal V}_{e\tau}^{12} D_1^{}  \vartheta_2^{} - {\cal J} D_2^{} \vartheta_1^{}\right) 
	\end{eqnarray}
	for $\alpha=e$ and $i=1$. We have replaced the light neutrino masses by $m_i^{} = D_i^{} t_i^{} \approx D_i^{} \vartheta_i^{}$. It indicates that the CP-violating asymmetries are not explicitly suppressed by the power of $\vartheta_i^{}$.
	
	In the scenario of {\it flavored resonant} leptogenesis, the expression of the baryon-to-photon ratio $\eta$ varies for different values of the equilibrium temperature $T \simeq M_i^{}$. In the so-called {\it $\tau$-flavored} leptogenesis with $10^9~{\rm GeV} \lesssim T \lesssim 10^{12}~{\rm GeV}$, the $\tau$-flavored Yukawa interaction will be in the thermal equilibrium and can be distinguished from the $e$ and $\mu$ flavor. The corresponding range for $\vartheta_1^{}$ with the chosen lightest neutrino mass is $10^{-12} \lesssim \vartheta_1^{} \lesssim 10^{-10}$. At this time, the baryon-to-photon ratio reads~\cite{Buchmuller:2002rq,Buchmuller:2003gz}
	\begin{eqnarray}
		\eta \simeq - 9.6 \times 10^{-3} \left[\sum_{\alpha = e,\mu} \left(\varepsilon_{1\alpha}^{} + \varepsilon_{2\alpha}^{}\right) \kappa \left(K_e^{} + K_\mu^{}\right) + \left(\varepsilon_{1\tau}^{} + \varepsilon_{2\tau}^{}\right) \kappa \left(K_\tau^{}\right) \right] \; .
	\end{eqnarray}
	Here the conversion efficiency factor $\kappa(K_\alpha^{})$ for $\alpha=e,\mu,\tau$ reflects the washout effects on the lepton number asymmetry by the inverse decays of heavy Majorana neutrinos and the lepton-number-violating scattering processes, which reads~\cite{Blanchet:2006be,Buchmuller:2004nz}
	\begin{eqnarray}
		\kappa (K_\alpha^{}) \equiv \frac{2}{K_\alpha^{} z_{\rm B}^{}(K_\alpha^{})} \left\{1-\exp\left[-\frac{1}{2} K_\alpha^{} z_{\rm B}^{}(K_\alpha^{})\right]\right\} 
	\end{eqnarray}
	with $z_{\rm B}^{} (K_\alpha^{}) \simeq 2 + 4 K_\alpha^{0.13} \exp\left(-2.5/K_\alpha^{}\right)$.	Meanwhile, the decay parameter $K_\alpha^{} \equiv K_{1\alpha}^{} + K_{2\alpha}^{}$ with $K_{i\alpha}^{} \equiv \widetilde{m}_{i\alpha}^{} / m_*^{}$ can be calculated from the equilibrium neutrino mass $m_*^{} \approx 1.07 \times 10^{-3}~{\rm eV}$~\cite{Blanchet:2006dq,Blanchet:2006be} and the effective light neutrino masses
	\begin{eqnarray}
		\widetilde{m}_{i\alpha}^{} \equiv \left|\left({\bf m}_{\rm D}^{}\right)_{\alpha i}^{}\right|^2 / M_i^{} \simeq I_\kappa^{-1} \left|\left({\bf u}_1^{}\right)_{\alpha i}^{}\right|^2 m_i^{} \left(1 - 2\Delta_\tau^{} \delta_{\alpha\tau}^{} \right) \;. 
	\end{eqnarray}
	When the equilibrium temperature decreases to $10^5~{\rm GeV} \lesssim T \lesssim 10^9~{\rm GeV}$, both the $\mu$- and $\tau$-flavored Yukawa interactions are in the thermal equilibrium, and contributions from three flavors in $\eta$ should be considered separately, with the baryon-to-photon ratio
	\begin{eqnarray}
		\eta \simeq - 9.6 \times 10^{-3} \sum_{e,\mu,\tau} \left(\varepsilon_{1\alpha}^{} + \varepsilon_{2\alpha}^{}\right) \kappa \left(K_\alpha^{}\right) \; .
	\end{eqnarray}
	In this {\it $(\mu+\tau)$-flavored} case, the angle $\vartheta_1^{}$ ranges between $10^{-10} \lesssim \vartheta_1^{} \lesssim 10^{-8}$. For higher temperatures, the relevant Yukawa interactions are blind to the lepton flavor, and there is no CP violation effect in such {\it unflavored} thermal leptogenesis.
	
	\begin{figure}[t]
		\centering
		\includegraphics[scale=0.9]{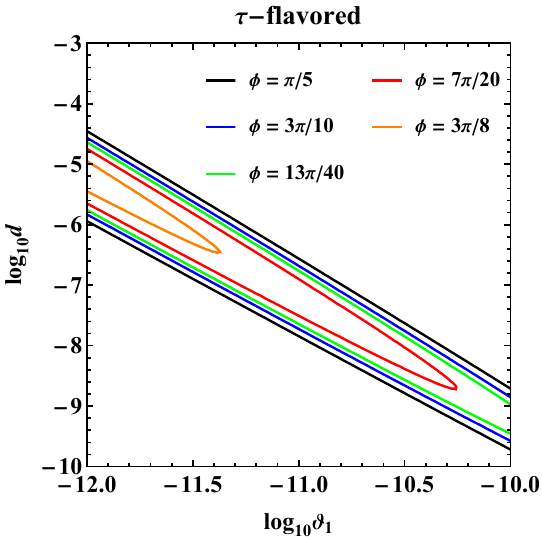}
		\includegraphics[scale=0.9]{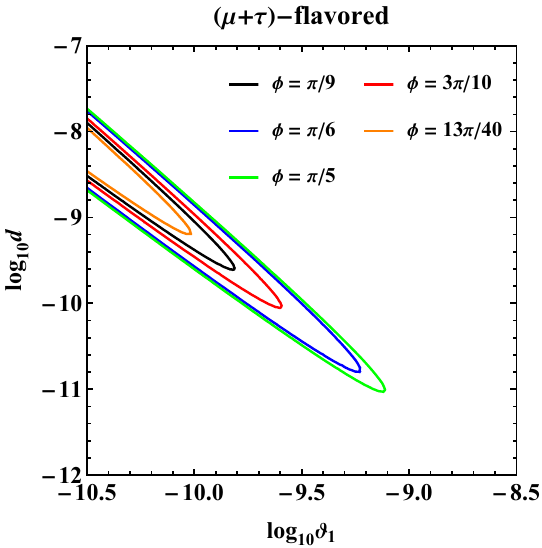}
		\caption{The parameter space of $\vartheta_1^{}$ and $d$ in the {\it flavored resonant} leptogenesis for the $\tau$-flavored regime (left panel) with $10^{-12} \lesssim \vartheta_1^{} \lesssim 10^{-10}$ and the $(\mu+\tau)$-flavored regime (right panel) with $10^{-10} \lesssim \vartheta_1^{} \lesssim 10^{-8}$ in the NO case. Five different values of $\phi$ are chosen as inputs to illustrate the contours. Other numerical values of physical parameters are the same as those in Sec.~\ref{sec:mutoegamma}.}
		\label{fig:eta}
	\end{figure}
	
	Using the same numerical values of input parameters as in Sec.~\ref{sec:mutoegamma}, we evaluate $\eta$ as the function of $\vartheta_1^{}$ and $\vartheta_2^{}$.\footnote{In flavored resonant leptogenesis $M_3^{}$ is not involved, so $\eta$ is independent on $\vartheta_3^{}$.} With the observed baryon-to-photon ratio $\eta \approx 6.04 \times 10^{-10}$~\cite{ParticleDataGroup:2024cfk}, we plot the corresponding parameter space in Fig.~\ref{fig:eta} for both the $\tau$-flavored (left panel) and the $(\mu+\tau)$-flavored (right panel) thermal leptogenesis in the NO case. As $M_1^{}$ and $M_2^{}$ are too close to each other, we define $d \equiv (M_2^{}-M_1^{})/M_1^{}$ to emphasize the difference between these two masses. With $M^{}_i = m^{}_i / t_i^2$, one may connect $d$ with $\vartheta_2^{}$ for given values of $\{\vartheta_1^{},m_1^{},m_2^{}\}$ with $t_2^{} = t_1^{}  \sqrt{m_2^{}} / \sqrt{m_1^{} (1 + d)}$. The contours represent the values of $\{\vartheta_1^{},d\}$, which satisfy the observed $\eta$, and five specific values of $\phi$ are selected for illustration. The parameter $d$ spans a range of $10^{-10} \lesssim d \lesssim 10^{-4}$ in the case of $\tau$-flavored leptogenesis, while in the $(\mu+\tau)$-flavored leptogenesis scenario, a narrower range of $d$ is observed with $10^{-11} \lesssim d \lesssim 10^{-8}$. Notably, the value of $d$ exhibits a negative correlation with $\vartheta_1^{}$, reflecting stronger degeneracy between $M_1^{}$ and $M_2^{}$. We do not find any parameter space in the IO case, since $m_1^{}$ and $m_2^{}$ are approximately an order of magnitude larger now, providing a larger $K_{i\alpha}^{}$ and depressing the conversion efficiency. It is clear that the flavored resonant leptogenesis, which generates the matter-antimatter asymmetry of the Universe, is able to narrow down the viable parameter regions and impose tighter constraints on $\vartheta_i^{}$ than those from LFV decays. 
	
	Although we focus on the strict and explicit calculations on physical observables in this work, the approximations from the RGEs running effects are inevitable. Specifically, when discussing the running behavior of neutrino masses, we adopt the $\overline{\rm MS}$ renormalization scheme and integrate out the heavy degrees of freedom to obtain the low-energy effective theory~\cite{Chankowski:1993tx,Babu:1993qv,Antusch:2001ck,Antusch:2005gp,Mei:2005qp,Ohlsson:2013xva}. This procedure itself involves certain approximations, such as considering only the running of the dimension-five operator and the corresponding Wilson coefficient $\kappa$, while neglecting higher-order contributions. A fully rigorous calculation requires retaining all degrees of freedom in the model, for which the on-shell renormalization scheme is the most suitable choice. In this case, the only uncertainty would come from the loop expansion, while higher-loop contributions can be safely neglected. A detailed study on the complete renormalization of the canonical seesaw model in the on-shell scheme and its phenomenological implications are left for future work.
	
	\section{Summary}
	
	\label{sec:sum}

	In this work, we perform an {\it explicit} calculation in the canonical seesaw model with the help of the Wigner-like parametrization, where the $6\times 6$ unitary matrix ${\cal U}$ is parametrized by four $3\times 3$ unitary matrices and three rotation angles $\left\{\vartheta_1^{}, \vartheta_2^{}, \vartheta_3^{}\right\}$. After redefining the lepton doublets and right-handed neutrino singlets in the flavor basis, the effective mass matrices $\widetilde{\bf m}_{\rm D}^{}$ and $\widetilde{\bf m}_{\rm R}^{}$ are both set to be diagonal, and the commutation relation $[{\bf m}_{\rm D}^\dagger {\bf m}_{\rm D}^{}, {\bf m}_{\rm R}^\dagger {\bf m}_{\rm R}^{}] = {\bf 0}$ holds. This special {\it ansatz} allows us to calculate physical observables explicitly. With the chosen physical parameters $\left\{m_i^{}, \theta_{ij}^{}, \delta, \rho, \sigma, \vartheta_i^{}\right\}$, we derive the LFV decay rates of charged leptons and evaluate the ratios of branching fractions $\xi_{\alpha\beta}^{}$. We also discuss the possibility of generating CP-violating asymmetries through flavored resonant leptogenesis with radiative corrections, and obtain the corresponding experimental constraints on $\vartheta_1^{}$ and the mass-difference $d$.
	
	Although the conclusions in this work come from special assumptions, demonstrating the phenomenological applications of the Wigner-like parametrization in such a case is still of great significance, and cannot bury the generality of parametrization itself. However, in the most general case, it remains infeasible to acquire exact analytical solutions expressed with physical parameters in the seesaw model. This comes from the intricate correspondence embedded in the seesaw relation. Specific choices of parametrization can partially alleviate these difficulties. Although different parametrization schemes are mathematically equivalent, the Wigner-like parametrization allows for an exact description of the seesaw relation in the mass basis and provides significant advantages in deriving any physical observables with fewer unknown physical parameters. Nevertheless, it is still challenging to ensure a fully self-consistent calculation at the one-loop level. 
	
	As neutrino physics enters the precision era, such explicit calculations without any approximation can help us constrain model parameters from experiments and understand the dynamic properties of the seesaw mechanism in a better way, such as the non-unitarity effects, the relations among original and derivative parameters, as well as the flavor mixing structure of active neutrinos at low energy. The connections between seesaw parameters and low-energy observables can also be established, which is crucial for future neutrino experiments. From this perspective, we hope that our work provides an example of calculations within a complete neutrino mass model for the whole field of neutrino physics.

	\section*{Acknowledgments}
	
	The author is indebted to Prof.~Zhi-zhong Xing and Prof.~Shun Zhou for carefully reading this manuscript and putting forward many useful suggestions. All Feynman diagrams are generated by {\tt FeynArts}~\cite{Hahn:2000kx}. This work was supported in part by the National Natural Science Foundation of China under grant No.~12475113, by the CAS Project for Young Scientists in Basic Research (YSBR-099), and by the Scientific and Technological Innovation Program of IHEP under grant No.~E55457U2.
	
	\appendix
	
	\section{Expressions of Jarlskog-like Invariants $\widetilde{\cal V}_{\alpha\beta}^{ij}$}
	\label{app:A}
	
	\setcounterpageref{equation}{0}
	
	\numberwithin{equation}{section}
	
	The Jarlskog-like invariant in the unitary case $\widetilde{\cal V}_{\alpha\beta}^{ij} \equiv {\rm Im}[\left({\bf u}^{}_1\right)_{\alpha i}^{} \left({\bf u}^{}_1\right)_{\beta i}^{} \left({\bf u}^{*}_1\right)_{\alpha j}^{} \left({\bf u}^{*}_1\right)_{\beta j}^{}]$ (for $\alpha,\beta=e,\mu,\tau$ and $i,j=1,2,3$) plays an important role in describing the neutrino-antineutrino oscillations and the CP-asymmetries in the thermal leptogenesis. From its definition, the following relations can be easily proved~\cite{Xing:2013woa}
	\begin{eqnarray}
		\label{eq:Vrelation}
		\widetilde{\cal V}_{e\mu}^{ij} &=& \frac{1}{2} \left(\widetilde{\cal V}_{\tau \tau}^{ij} - \widetilde{\cal V}_{ee}^{ij} - \widetilde{\cal V}_{\mu\mu}^{ij}\right) \;, \nonumber \\
		\widetilde{\cal V}_{e\tau}^{ij} &=& \frac{1}{2} \left(\widetilde{\cal V}_{\mu\mu}^{ij} - \widetilde{\cal V}_{ee}^{ij} - \widetilde{\cal V}_{\tau\tau}^{ij}\right) \;, \nonumber \\
		\widetilde{\cal V}_{\mu\tau}^{ij} &=& \frac{1}{2} \left(\widetilde{\cal V}_{ee}^{ij} - \widetilde{\cal V}_{\mu\mu}^{ij} - \widetilde{\cal V}_{\tau\tau}^{ij}\right) \;,
	\end{eqnarray}
	together with $\widetilde{\cal V}_{\alpha\beta}^{ij} = \widetilde{\cal V}_{\beta\alpha}^{ij} = - \widetilde{\cal V}_{\alpha\beta}^{ji} = - \widetilde{\cal V}_{\beta\alpha}^{ji}$. Therefore, only nine invariants are independent, which are usually chosen as $\widetilde{\cal V}_{\alpha\alpha}^{ij}$ (for $\alpha=e,\mu,\tau$ and $ij=12,13,23$). Their specific expressions in the standard parametrization of ${\bf u}_1^{}$ can be written as
	\begin{eqnarray}
		\label{eq:Vee}
		\widetilde{\cal V}_{ee}^{12} &=& s_{12}^2 c_{12}^2 c_{13}^4 \sin 2(\rho-\sigma) \;, \nonumber\\
		\widetilde{\cal V}_{ee}^{13} &=& c_{12}^2 s_{13}^2 c_{13}^2 \sin 2(\delta+\rho) \;, \nonumber\\
		\widetilde{\cal V}_{ee}^{23} &=& s_{12}^2 s_{13}^2 c_{13}^2 \sin 2(\delta+\sigma)
	\end{eqnarray}
	for $\alpha=e$, and 
	\begin{eqnarray}
		\label{eq:Vmumu}
		\widetilde{\cal V}_{\mu\mu}^{12} &=& s_{12}^2 c_{12}^2 \left(c_{23}^4-4 s_{13}^2 s_{23}^2 c_{23}^2 +s_{13}^4 s_{23}^4\right) \sin 2(\rho-\sigma)  \nonumber \\
		&& +2 s^{}_{12} c_{12}^{} s^{}_{13} s^{}_{23} c^{}_{23} \left(c_{23}^2-s_{13}^2 s_{23}^2\right) \left[c_{12}^2 \sin (2 \rho-2 \sigma+\delta)-s_{12}^2 \sin (2 \rho-2 \sigma-\delta)\right] \nonumber \\
		&& +s_{13}^2 s_{23}^2c_{23}^2 \left[s_{12}^4 \sin 2(\rho-\sigma-\delta) + c_{12}^4 \sin 2(\rho-\sigma+\delta)\right] \;, \nonumber\\
		\widetilde{\cal V}_{\mu\mu}^{13} &=& c_{13}^2 s_{23}^2\left[s_{12}^2 c_{23}^2 \sin 2 \rho + 2 s^{}_{12} c^{}_{12} s^{}_{13} s^{}_{23} c^{}_{23} \sin (\delta+2 \rho) + c_{12}^2 s_{13}^2 s_{23}^2 \sin 2(\delta+\rho)\right] \;, \nonumber\\
		\widetilde{\cal V}_{\mu\mu}^{23} &=& c_{13}^2 s_{23}^2 \left[c_{12}^2 c_{23}^2 \sin 2 \sigma - 2 s^{}_{12} c^{}_{12} s^{}_{13} s^{}_{23} c^{}_{23} \sin (\delta+2 \sigma) + s_{12}^2 s_{13}^2 s_{23}^2 \sin 2(\delta+\sigma)\right]
	\end{eqnarray}
	for $\alpha=\mu$, and 
	\begin{eqnarray}
		\label{eq:Vtautau}
		\widetilde{\cal V}_{\tau\tau}^{12} &=& s_{12}^2 c_{12}^2 \left(s_{23}^4-4 s_{13}^2 s_{23}^2 c_{23}^2 + s_{13}^4 c_{23}^4\right) \sin 2(\rho-\sigma) \nonumber \\
		&& -2 s^{}_{12} c^{}_{12} s^{}_{13} s^{}_{23} c^{}_{23} \left(s_{23}^2-s_{13}^2 c_{23}^2\right)\left[c_{12}^2 \sin (2 \rho-2 \sigma+\delta)-s_{12}^2 \sin (2 \rho-2 \sigma-\delta)\right] \nonumber \\
		&& + s_{13}^2 s_{23}^2 c_{23}^2 \left[s_{12}^4 \sin 2(\rho-\sigma-\delta) + c_{12}^4 \sin 2(\rho-\sigma+\delta)\right] \;, \nonumber\\
		\widetilde{\cal V}_{\tau\tau}^{13} &=& c_{13}^2 c_{23}^2\left[s_{12}^2 s_{23}^2 \sin 2 \rho - 2 s^{}_{12} c^{}_{12} s^{}_{13} s^{}_{23} c^{}_{23} \sin (\delta+2 \rho) + c_{12}^2 s_{13}^2 c_{23}^2 \sin 2(\delta+\rho)\right] \;, \nonumber\\
		\widetilde{\cal V}_{\tau\tau}^{23} &=& c_{13}^2 c_{23}^2\left[c_{12}^2 s_{23}^2 \sin 2 \sigma + 2 s^{}_{12} c^{}_{12} s^{}_{13} s^{}_{23} c^{}_{23} \sin (\delta+2 \sigma) + s_{12}^2 s_{13}^2 c_{23}^2 \sin 2(\delta+\sigma)\right] 
	\end{eqnarray}
	for $\alpha = \tau$. Note that only invariants with $ij=12$ appear in Eq.~(\ref{eq:epsilon-resonant}) and all depend on $\phi \equiv \rho-\sigma$.

\end{document}